\documentclass[11pt,nofootinbib]{revtex4}
\usepackage{amsfonts,amssymb,amsmath,graphicx}
\def\dac{\displaystyle\frac}

\def\[{\left[}
\def\]{\right]}
\def\({\left(}
\def\){\right)}

\newcommand{\diag}{\mathop{\rm diag}\nolimits}

\begin{document}

\baselineskip7mm
\title{The nature of singularity in multidimensional anisotropic
Gauss-Bonnet cosmology with a perfect fluid}

\author{I.V. Kirnos}
\affiliation{Tomsk State University, Tomsk, 634050 Russia}

\author{A.N. Makarenko}
\affiliation{Tomsk State Pedagogical University, Tomsk, 634041
Russia}

\author{S.A. Pavluchenko}
\affiliation{Special Astrophysical Observatory, Russian Academy of
Sciences, Nizhnij Arkhyz, 369167 Russia}

\author{A.V. Toporensky}
\affiliation{Sternberg Astronomical Institute, Moscow State
University, Moscow, 119992 Russia}

\begin{abstract}
We investigate dynamics of (4+1) and (5+1) dimensional flat
anisotropic Universe filled by a perfect fluid in the Gauss-Bonnet
gravity. An analytical solutions valid for particular values
of the equation of state parameter $w=1/3$ have been found. For other
values of $w$ structure of cosmological singularity have been
studied numerically. We found that for $w > 1/3$ the singularity
is isotropic. Several important differences between (4+1) and
(5+1) dimensional cases are discussed.
\end{abstract}

\maketitle

\section{Introduction}

Recent several years show increasing interest to modified gravity
both for $(3+1)$-dimensional Universe (which is mostly motivated
by attempts to explain observed accelerated expansion of our
Universe) and for multidimensional cosmological models which are
more speculative, though they pose important questions on the
nature of cosmological singularity and possible evolution of the Universe
 in its
very early epoch. In the latter
perspective the Lovelock gravity \cite{Lovelock} is one of the very popular
extension of General Relativity (GR)
(see, for example, \cite{N1,Kitaura,moreGB,Maeda1,Maeda2,Maeda3,Ferraro,Barrau,Maeda4,
Cai,Tomsk,Giribet,D1,D2}). It keeps the order of corresponding
equations of motion unchanged with respect to GR, and , as it was
recently claimed in \cite{Paddy}, nice thermodynamical properties
of GR survive in Lovelock gravity in contrast to, for example,
$f(R)$ theories.

In the present paper we consider $(4+1)$ and $(5+1)$ dimensional
anisotropic Universe. In these dimensions the only non-Einstein
term in the Lovelock action is the famous Gauss-Bonnet
combination. Power-law solutions which replace Kasner regime in
the Gauss-Bonnet gravity in a flat anisotropic multidimensional
Universe have been studied since the end of 80s \cite{Deruelle}.
However, all previous works have been devoted to vacuum solutions.
Importance of the Kasner solution (which is a vacuum anisotropic
solution in GR) follows from the fact that an ordinary barotropic
fluid different from a massive scalar field is dynamically
unimportant near a cosmological singularity. It means that we can still
use the Kasner solution to describe initial singularity in a
Universe filled by rather general kind of an ordinary matter. The
goal of the present paper is to provide similar analysis for
multidimensional Universe in the Lovelock gravity.

\section{Main equations}

We consider a flat anisotropic metric in (n+1)-dimensional
space-time. We are dealing with Einstein-Gauss-Bonnet gravity, and
non-vacuum space-time. Lagrangian of this theory have a form

$$
{\cal L} = R + \alpha{\cal L}_2 + {\cal L}_M,
$$

\noindent where $R$ is Ricci scalar, ${\cal L}_M$ is the
Lagrangian of matter fields and ${\cal L}_2$

\begin{equation}
{\cal L}_2 = R_{\mu \nu \alpha \beta} R^{\nu \mu \alpha \beta} - 4
R_{\mu \nu} R^{\mu \nu} + R^2 \label{lagr1}
\end{equation}

\noindent is the Gauss-Bonnet Lagrangian.

In the present paper we are interesting in the behavior mostly in
the vicinity of the cosmological singularity, which allows us to
take into account only corrections of the highest possible order.
In our case it is the Gauss-Bonnet contribution,
so we neglect Einstein terms. In the absence of matter sources this
problem have been studied in \cite{Deruelle, TT}, in the present paper
we take matter into account. The full Einstein - Gauss-Bonnet
system shows a complicated behavior even in the vacuum case \cite{In,PT},
and we leave investigation of such system with matter for a future work.


We are working in the flat background, so the metric we
considering has a form

\begin{equation}\label{metric2}
g_{\mu\nu} = \diag\{ -1, a_1^2(t), a_2^2(t),\ldots, a_n^2(t)\}.
\end{equation}

We use perfect fluid with the equation of state $p=w\rho$ as a
matter source;
after varying action obtained from~(\ref{lagr1}) using the metric
above and perfect fluid as a matter field, one can obtain the
following equation of motion: there are $n$ dynamical equations and a constraint
equation. The $j$-th dynamical equation has the form

\begin{equation}\label{eqmotion}
-3\sum_{\substack{i,k,l,m\neq j\\
i<k<l<m}}H_i H_k H_l H_m-\sum_{i\neq j}(H_i^2 + \dot
H_i)\sum_{\substack{k,l\neq j,i\\ k<l}}H_k H_l=w\sigma\rho;
\end{equation}

\noindent while constraint equation becomes

\begin{equation}\label{constr}
3\sum_{i<j<k<l}H_i H_j H_k H_l=\sigma\rho,
\end{equation}

\noindent
where $\sigma=
2\pi G/\alpha c^4$ is a combination of fundamental constants.
 We assume $\alpha > 0$ in order to
avoid instabilities in quantum theory. Also
we need the continuity equation:

\begin{equation}\label{cont_eq}
\dot \rho + (\rho + p) \sum\limits_{i=1}^n H_i = 0.
\end{equation}

To demonstrate the difference between (4+1) and (5+1) cases let us
write down respective equation separately for these two cases. For
(4+1) case we have

\begin{equation}\label{eqmot4}
(\dot H_b+H_b^2)H_cH_d +(\dot H_c+H_c^2)H_bH_d +(\dot
H_d+H_d^2)H_bH_c + w\sigma\rho =0
\end{equation}

\noindent
 as a dynamical equation (the rest three can be obtained
by cyclic transmutation of indices) and the following constraint:

\begin{equation}\label{const4}
3 H_aH_bH_cH_d = \sigma\rho.
\end{equation}

For (5+1) case we have

\begin{equation}\label{eqmot5}
\begin{array}{l}
3H_bH_cH_dH_f+ (\dot H_b+H_b^2)(H_cH_d+H_cH_f+H_dH_f) +(\dot H_c+H_c^2)(H_bH_d+H_bH_f+H_dH_f) \\
\\ +(\dot H_d+H_d^2)(H_bH_c+H_bH_f+H_cH_f)+(\dot
H_f+H_f^2)(H_bH_c+H_bH_d+H_cH_d) +
w\sigma\rho =0
\end{array}
\end{equation}

\noindent
 as dynamical equations (rest four can be obtained the
same way as for (4+1) case) and

\begin{equation}\label{constr5}
3
\[ H_aH_bH_cH_d+H_aH_bH_cH_f+H_aH_bH_dH_f+H_aH_cH_dH_f+H_bH_cH_dH_f \]
= \sigma\rho
\end{equation}

\noindent
 as the constraint.

 Now one can clearly see the differences
between these two cases. First of all, the structure of the
dynamical equations in the (4+1) case (\ref{eqmot4}) lacks the
term with the product of four Hubble parameters. Indeed, this term
in the $j$-th equation consists of the product of four Hubble
parameters with indices different from $j$ and each other. In
(4+1) dimensions we have only three indices different from a given
$j$, so this term is absent. In the (5+1) case (\ref{eqmot5})
there exits exactly four such indices, so one possible term
exists.
 In higher dimensions
with a bigger number of  Hubble parameters the corresponding sum has
more than one term.
The second important difference exists in the
structure of the constraint equations -- in the (4+1) case the
left-hand side of
(\ref{const4}) has only one term (all possible sums of four
different Hubble parameters needed) while in the (5+1) case (\ref{constr5})
it has five of them. Both these features result in interesting
differences in dynamical behavior of (4+1)- and (5+1)-dimensional
models (see below).


\section{Power-law solutions}

Now let us find some exact solutions with one particular {\it
anzatz}. These solutions are of power-law-type, so scale factors
take a form $a_i(t) = t^{p_i}$, after substituting Hubble
functions $H_i = p_i/t$ and their derivatives $\dot H_i = -
p_i/t^2$ into~(\ref{eqmotion}) and~(\ref{constr}) one can rewrite
constraint equation



\begin{equation}3\sum_{i<j<k<l}p_i p_j p_k
p_l=\sigma\rho_0\, t^{4-(1+w)\sum_i p_i},
\end{equation}

\noindent as well as equations of motion

\begin{equation}-3\sum_{\substack{i,k,l,m\neq j\\
i<k<l<m}}p_i p_k p_l p_m-\sum_{i\neq
j}p_i(p_i-1)\sum_{\substack{k,l\neq j,i\\ k<l}}p_k p_l=w\sigma\rho_0\,
t^{4-(1+w)\sum_i p_i};
\end{equation}

\noindent there $\rho_0$ is the matter density at some given
moment of time.

It is clear that if $\rho\neq 0$ then these equations can be
solved only under condition
\begin{equation}\label{condition}\sum_i p_i=\dac{4}{1+w}.\end{equation}

Hence, field equations are
\begin{equation}\label{00-equation}3\sum_{i<j<k<l}p_i p_j p_k
p_l=\sigma\rho_0,\end{equation}
\begin{equation}\label{jj-equation}-3\sum_{\substack{i,k,l,m\neq j\\
i<k<l<m}}p_i p_k p_l p_m-\sum_{i\neq
j}p_i(p_i-1)\sum_{\substack{k,l\neq j,i\\ k<l}}p_k
p_l=w\sigma\rho_0.\end{equation}

In Einstein gravity Kasner solution (for vacuum) implies $\sum_i
p_i=1$, and such a condition conserves in Jacobs solution (for
maximally stiff fluid) (for Kasner and Jacobs solutions see e.~g.
\cite{Jacobs,DSC}). Assume now that similarly in Gauss-Bonnet
gravity condition $\sum_i p_i=3$ (obtained earlier for vacuum)
conserves for some kind of matter. One should keep in mind, though,
that for vacuum (4+1) case we still have $\sum_i p_i=3$ but under
condition that one of $p_i$ is always zero -- indeed, using {\it
anzatz} above to Eq.\,(\ref{const4}) one can see that one of $p_i$
should be always zero.

From $\sum_i p_i=3$ and (\ref{condition}) one can obtain $w=1/3$.
Translate now equations (\ref{00-equation}) and (\ref{jj-equation})
while taking $w=1/3$ and $\sum_i p_i=3$ into account. For example,
\begin{equation}\label{example}\begin{array}{l}\displaystyle 24\sum_{i<j<k<l}p_i p_j p_k p_l=\sum_i p_i
\sum_{j\neq i} p_j \sum_{k\neq i,j} p_k \sum_{l\neq i,j,k}p_l=\\
\quad{}\displaystyle =\sum_i
p_i \sum_{j\neq i} p_j \sum_{k\neq i,j} p_k (3-p_i-p_j-p_k)=\vphantom{\(\sum_i p_i^2\)^2}\\
\quad{}\displaystyle =3\sum_i p_i\sum_{j\neq i} p_j \sum_{k\neq
i,j}
p_k-3\sum_i p_i^2\sum_{j\neq i} p_j \sum_{k\neq i,j} p_k=\ldots=\vphantom{\(\sum_i p_i^2\)^2}\\
\quad{}\displaystyle =81-54\sum_i p_i^2+24\sum_i p_i^3-6\sum_i
p_i^4+\(\sum_i p_i^2\)^2.\end{array}\end{equation} At that all
equations (\ref{00-equation}), (\ref{jj-equation}) turns out to be
identical with each other. Any of them takes a form
\begin{equation}81-54\sum_i p_i^2+24\sum_i p_i^3-6\sum_i
p_i^4+\(\sum_i p_i^2\)^2=8\sigma\rho_0.\end{equation} Therefore, with
$w=1/3$ considered metric with pointed above {\it anzatz} will be
an exact solution only if
\begin{equation}\label{power-law-any-dimension}\sum_i
p_i=\dac{4}{1+w}=3,\qquad 3\sum_{i<j<k<l}p_i p_j p_k
p_l=\sigma\rho_0.\end{equation}

Cases $n=4$ and $n=5$ were investigated for arbitrary $w$. At
$n=4$ there are no other anisotropic power-law solutions with
nonzero energy density, but for $n=5$ there is the following
solution:
$$p_1=p_2=p_3\equiv p,\qquad p_4=p_5=-\dac{1}{2}p,$$
$$w=-\dac{p-2}{p},\qquad \sigma\rho_0=-\dac{3}{4}p^4.$$ This solution have
a remarkable feature: if 3 visible dimensions expanse isotropically
then extra dimensions contract isotropically. However, here
$\sigma\rho_0<0$, therefore, since $\sigma = \frac{2\pi G}{\alpha
c^4}$ either $\rho_0<0$, or $\alpha<0$. The latter means no global
energy minimum and hence instability in quantum theory. So this
solution seems to have no physical meaning; we wrote it down here
just for mathematical completeness.

\section{Numerical analysis}

In order to understand what happens for $w \ne 1/3$ we remind a
reader similar situation in the Einstein gravity. Kasner solution
gives us a condition $\sum_i p_i = 1$ which indicates that the
volume expands linearly with time, and, correspondingly, the
matter density in the vicinity of Kasner regime decreases as $\rho
\sim V^{-w-1} \sim t^{-w-1}$. On the other hand, left-hand side of
equations of motion decreases as $t^{-2}$. These two scaling rates
coincide at $w=1$ (the Jacobs solution). For $w < 1$ a small
amount of matter becomes dynamically unimportant for evolution
towards a cosmological singularity, and ultimately disturbs the
Kasner solution for an expanding Universe.

Of course, in the Einstein gravity we know the situation much better,
because the dynamical equations describing Bianchi-I Universe with
a perfect fluid can be solved exactly. This solution tells us that
the Kasner regime is the unique past attractor in the case $w < 1$
independently of the initial conditions, while isotropic regime is
the unique future attractor \cite{DSC}. Though $w$ can not exceed the value
of $1$ for any stable ``physical'' matter, there are some
situation (for example in brane cosmology \cite{Deffayet}) where the dynamics can
be described as it would be driven by a matter with an effective
$w > 1$, and in this case the picture is the opposite one with
past isotropic and future Kasner attractors \cite{Varun,T}.


\begin{figure}[t]
\includegraphics[width=1.0\textwidth,bb=0 0 647 647,clip]{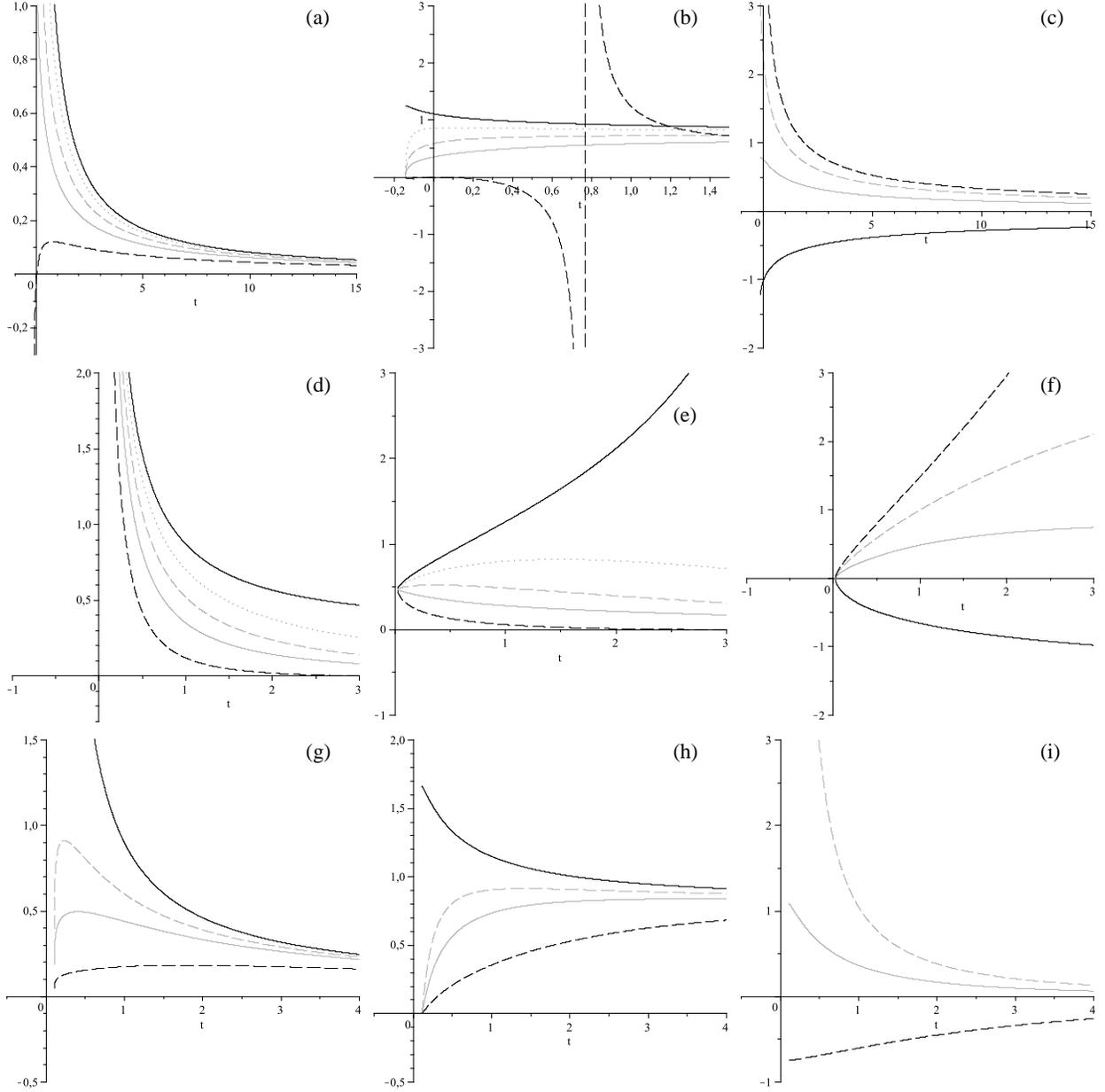}
\caption{Behavior of individual Hubble parameters (a, d, g),
Kasner exponents (b, e, h) and relative Hubble differences (c, f,
i) for $\[\mbox{(5+1)},~w< 1/3\]$ (first row), $\[\mbox{(5+1)},~w>
1/3\]$ (middle row) and $\[\mbox{(4+1)},~w< 1/3\]$ (last row)
cases. Note that $x$-axis $t$ is not a cosmological time, but rather
an internal parameter of our numerical calculations. This implies for both Fig. 1
and Fig. 2.}
\end{figure}

\begin{figure}[t]
\includegraphics[width=1.0\textwidth,bb=0 0 647 431,clip]{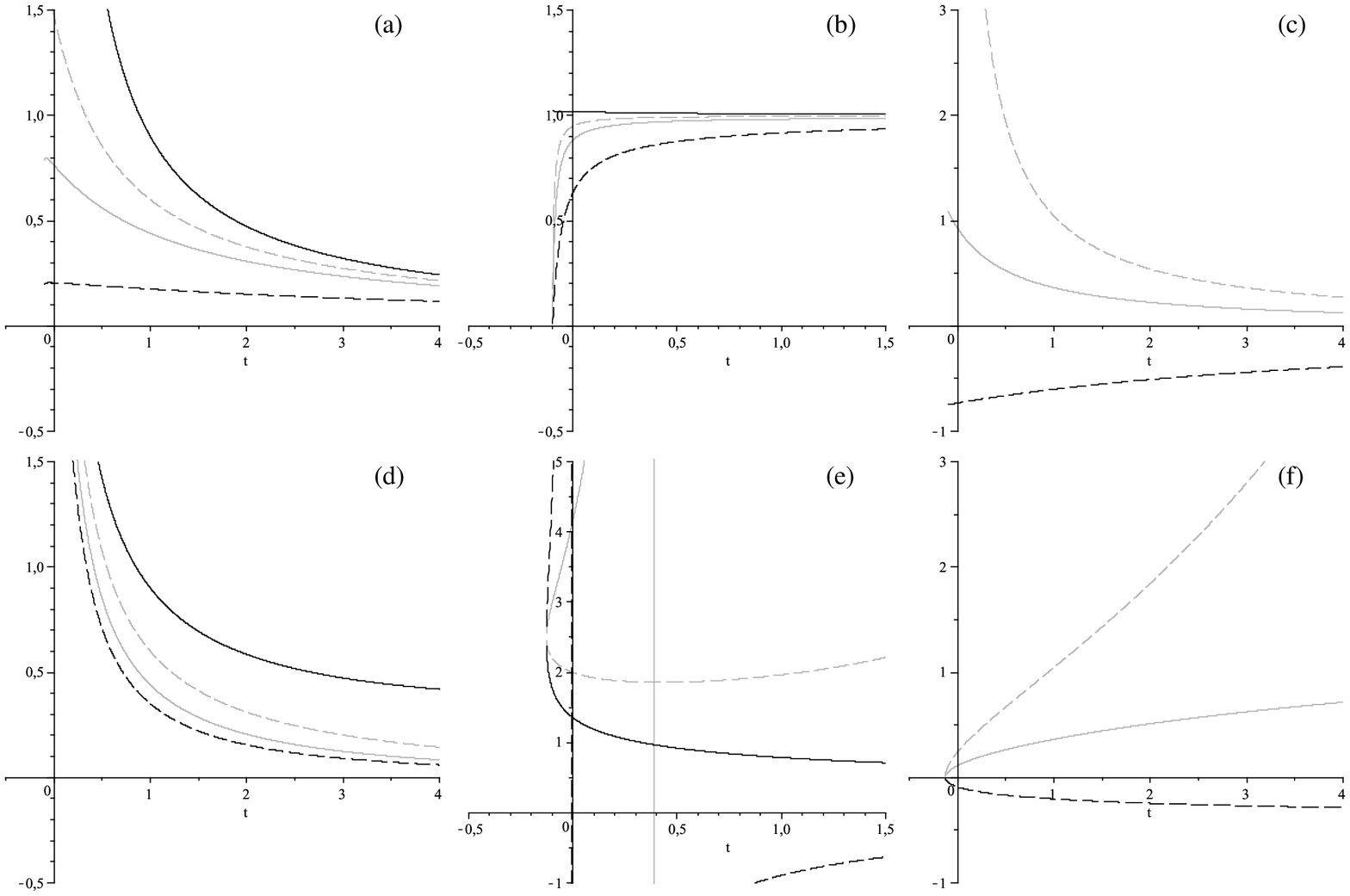}
\caption{Behavior of individual Hubble parameters (a, d), Kasner
exponents (b, e) and relative Hubble differences (c, f) for
$\[\mbox{(4+1)},~w=-0.01\]$ (first row), $\[\mbox{(4+1)},~w>
1/3\]$ (last row) cases.}
\end{figure}

In the Gauss-Bonnet gravity the situation is as follows. Left-hand
 sides of Eqs. (3, 4) scales as $t^{-4}$ in vacuum, while volume
increases as $t^3$. Adding a matter we obtain $\rho \sim V^{-w-1}
\sim t^{-3w-3}$ if the density of matter is small enough to keep
the volume expansion rate close to its vacuum value. Comparing
these two rates we see that the power-law vacuum solution is
destroyed by matter if $w > 1/3$ for a contracting Universe and
for $w < 1/3$ during expansion. On the other hand, the power-law
solution is stable with respect to adding a small amount of matter
with $w < 1/3$ for contracting and with $w > 1/3$ for expanding
Universe.

In order to study global stability of the power-law solution and
check whether its instability for appropriate values of $w$ leads
to isotropisation we provide numerical analysis of the system (3,
4) for various initial conditions and $w$. The results we have
found are different for $(4+1)$- and $(5+1)$-dimensions.

In the $(5+1)$-dimensional case the situation is similar to the
General Relativity dynamics described above. In Fig.\,1\,(a, b, c)
we plot typical behavior of power indices for $w < 1/3$. There we
have individual Hubble parameters in (a) panel, individual Kasner
exponents in (b) and relative differences ($H_i - H_1$ where $i\ne
1$) in (c) panel. We have found that for all studied initial
conditions the past singularity is anisotropic and described by
above mentioned power-law solution. And although in Fig.\,1\,(b)
four Kasner exponents tends to the same value (which is equal to
zero), from Fig.\,1\,(c) one can verify that the singularity is
anisotropic -- differences are non-zero and not equal to each
other.  On the other hand, the singularity is isotropic (again,
independently of initial conditions) if $w > 1/3$ (see
Fig.\,1\,(d, e, f)). In this case one can see that all Kasner
exponents tend to the same and non-zero value (Fig.\,1\,(e)) and
Hubble differences all tend to zero (Fig.\,1\,(f)).

The situation in $(4+1)$ dimensions is more interesting. First of
all, Eq. (6) indicates that while in the vacuum case at least one
Hubble parameter should vanish, even a small amount of matter
makes solutions with vanishing Hubble parameters impossible. This
means that an arbitrary small $\rho$ changes properties of the
solution significantly and can not be treated as a perturbation at
any given moment of time. Thus, dynamics of $(4+1)$ dimensional
Bianchi-I Universe can show features completely different from
those in the vacuum case.

We have found such qualitative difference in case of pressureless
matter $w=0$. It is easy to see that the conditions
$p_1=p_2=p_3=p_4=1$ satisfy the system (5, 6) (we use $p_i = -
H_i^2 / \dot H_i$), providing an exact solution, which describes
anisotropic Universe with constant (given by initial data and
not changing with time) ratio of Hubble parameters. Note that for
a higher-dimensional Universe a product of four Hubble parameters
in the Eq.(5) destroys this solution, so it is valid only in (4+1)
dimensions.

Typical behavior of power indices for other $w < 1/3$ is shown in
Fig.\,1\,(g, h, i). It is quite similar to the $(5+1)$ case with
$w < 1/3$, but there are some differences, like tending some
Hubble parameters to zero.
In theories with Einstein term in the action this feature can violate
dominance of Gauss-Bonnet contribution near a singularity, so this case
needs more work.
Initial singularity appears to be
anisotropic with at least one of the Hubble parameters tending to
zero. The last property is absent for $w=0$, and this value is an
exceptional one: any nonzero $w$ leads to a qualitatively similar
pictures, as it can be seen in Fig.\,2\,(a, b, c) plotted for
$w=-0.01$. Since $w=-0.01$ satisfy $w < 1/3$ condition, it looks
the same with ``usual'' $\[\mbox{(4+1)},~w< 1/3\]$ case. With
$w\to 0$ the sudden drop of Kasner exponents to zero becomes
sharper and sharper, while in limit $w=0$ it disappears, and all
Kasner exponents become equal to unity all the time, as explained
above.

On the other hand, we have found that independently of initial
conditions and the equation of state parameter, the initial
singularity for the case of $w>1/3$ is isotropic (see Fig.\,2\,(d,
e, f)). The situation is the same as for $\[\mbox{(5+1)},~w>
1/3\]$ case.

\section{Conclusions}

We have studied the cosmological dynamics of a flat anisotropic
multidimensional universe filled with a barotropic fluid in the
Gauss-Bonnet gravity. Power-law analytical solutions existing for
particular value of the equation of state parameter ($w=1/3$)
which are analogous to the Jacobs solution in the General
Relativity existing for $w=1$ have been found.

In the $(5+1)$-dimensional case this analogy exists also for other
values of $w$, leading to isotropic nature of the initial
singularity for $w > 1/3$ and anisotropic singularity for $w <
1/3$. In $(4+1)$ dimensions the picture is less clear. First of
all, an exceptional solution with constant anisotropy exists for
$w=1$. Second, though our numerical results show clearly that
singularity for $w < 1/3$ is anisotropic, some Hubble parameters
tend to zero, indicating that neglected Einstein term in the
action may be important in this case. This issue requires special
investigation. In case of $w > 1/3$ singularity is isotropic in
$(4+1)$ as well as in $(5+1)$ dimensions.

\section*{Acknowledgments}
This work is partially supported by RFBR grant 08-02-00923 and
scientific school grants 2553.2008.2 and 4899.2008.2.

\end{document}